\documentclass[12pt]{article}
\usepackage[dvips]{epsfig}
\usepackage{latexsym}
\usepackage[bf,footnotesize]{caption}	% Customize captions
\begin{document}
\setlength{\unitlength}{1mm}

\newcommand{\ba} {\begin{eqnarray}}
\newcommand{\ea} {\end{eqnarray}}
\newcommand{\be}{\begin{equation}}
\newcommand{\ee}{\end{equation}}
\newcommand{\n}[1]{\label{#1}}
\newcommand{\eq}[1]{Eq.(\ref{#1})}
\newcommand{\ind}[1]{\mbox{\tiny{#1}}}
\renewcommand\theequation{\thesection.\arabic{equation}}

\newcommand{\la}{\langle}
\newcommand{\ra}{\rangle}

\newcommand{\nn}{\nonumber \\ \nonumber \\}
\newcommand{\nl}{\\  \nonumber \\}
\newcommand{\pr}{\partial}
\renewcommand{\vec}[1]{\mbox{\boldmath$#1$}}

\title{{\hfill {\small Alberta-Thy-04-00 } } \vspace*{2cm} \\
Quantum Radiation of a Uniformly Accelerated Refractive Body}
\author{\\
V. Frolov\thanks{e-mail: frolov@phys.ualberta.ca}
\  and 
D. Singh\thanks{e-mail: 
singh@phys.ualberta.ca}
\date{\today}}
\maketitle
\noindent 
{
\centerline{ \em
Theoretical Physics Institute, Department of Physics,} \\ 
\centerline{ \em University of Alberta, Edmonton, Canada T6G 2J1}
}
\bigskip

\begin{abstract}
We study quantum radiation generated by an accelerated motion of a
small body with a refractive index $n$ which differes slightly from 1.
To simplify calculations we consider a model with a scalar massless
field. We use  the perturbation expansion in a small parameter $n-1$ to
obtain a correction to the vacuum Hadamard function for a uniformly
accelerated motion of the body. We obtain the vacuum expectation
for the energy density flux in the wave zone and discuss its
properties.
\end{abstract}

\bigskip

\centerline{\it PACS number(s): 03.70.+k, 11.10.-z, 42.50.Lc}

\baselineskip=.6cm

\newpage

\section{Introduction} 

Electromagnetic radiation from a uniformly accelerated charge is an
example of an ``eternal problem'' of physics. Starting with the work by
Born \cite{Born:09} different aspects of this problem have been
discussed again and again (see e.g \cite{FuRo:60,Ginz:69} and
references therin). In this paper we consider a peculiar quantum
analogue of this problem. 

Consider a small uncharged body and assume that it has internal degrees
of freedom interacting with the electromagnetic field. A polarizable
body is a well known example of such a system.  Electromagnetic
zero-point fluctuations induce dipole  moments in the body.  If a
polarizable body is at rest,  corrections to the field connected with
this  effect after averaging result in the change of the energy of the
system. In the presence of two polarizable bodies the energy shift
depends on the distance between them and results in the well known
Casimir effect (see e.g. \cite{Casi:48}--\cite{CASI}).  For an
accelerated motion of a polarizable body the net result of the emission
of its induced fluctuating dipole moment is quantum radiation, or so
called dynamical Casimir effect  \cite{DeWi:75}--\cite{BiDa:82}.   For
a non-relativistic motion  and relativistic motion in 2 dimensional
spacetime this effect is studied  quite well, especially in the special
case when the polarizability is very high and a surface of the body can
be approximated by a reflecting mirror-like boundary
\cite{Moor:70}--\cite{BaNo:96}.  Much less results have been obtained
for the relativistic motion in physical 4 dimensional spacetime.  
Exception are cases of a uniformely accelerated plane mirror
\cite{CaDe:77} and relativistic expanding spherical mirrors
\cite{FrSe:80}--\cite{FrSi:99}. See also \cite{KuFr:87} where quantum
radiation from an accelerated spherical body with a mirror boundary was
considered. In this paper we study the effect of quantum radiation
generated by an accelerated motion of a small polarizable body. 

To simplify calculations,  we assume that  internal degrees of freedom
of a body interact not with the electromagnetic field but with  a
scalar massless field. We assume that a ``refractive index'' $n$ inside
a small ball differs from its vacuum value, 1.   To solve the problem,
we assume also that $n$ is close to 1 and use  the perturbation
expansion in a small parameter $n-1$. We consider a simplest
accelerated motion when the direction and the value of the acceleration
(as measured in the comoving frame) are constant. We calculate a
correction to the  vacuum Hadamard function created by a uniformly
accelerated motion  of the body. The main result of the paper is the
expression for the quantum average energy density flux at infinity for
this problem. 

The paper is organized as follows. Section 2 discusses the set up
of the problem. We formulate the equation for a massless scalar field
in a presence of ``refracting'' matter and discuss the case when a
refractive ball is uniformely accelerated.  Section 3 contains  the
calculation of the perturbed Hadamard function in the presence of the
accelerated refractive ball of small size. We also discuss symmetry
relations for observables at ${\cal J}^+$ connected with the boost
invariance of the problem. The expressions for $\langle
\hat{\varphi}^2\rangle^{\ind{ren}}$ and $\langle
\hat{T}^{\mu}_{\nu}\rangle^{\ind{ren}}$ in the wave zone  are obtained
and discussed in Section~4.  Section~5 contains a discussion of the
obtained results.

\section{Model}\label{s2}
\setcounter{equation}0
\subsection{A ``refractive''  body in a static spacetime}

Our purpose is to study quantum radiation from a  an accelerated
``refractive'' body. We shall use a simplified model by  assuming that
a body interacts with a quantum massless scalar field $\varphi$.

In order to describe the model let us consider first a static gravitational
field described by metric
\be
dS^2 = -A^2\,dt^2 + \gamma_{ij}\,dx^i\,dx^j\,. \n{2.1}
\ee
Here $A$ and $\gamma_{ij}$ do not depend on $t$; $\xi^{\mu}\,
\partial_{\mu}=\partial_{t}$ is a Killing vector, and
\be
A^2=-\xi^{\mu}\,\xi_{\mu}\,. \n{2.2}
\ee

It is well known (see \cite{LaLi:62}) that the Maxwell equations
in the metric (\ref{2.1}) are identical to the Maxwell equations
in a media with $\varepsilon = \mu = A$. Thus $A^{-1}$ plays the role of 
the refraction index $n=1/\sqrt{\varepsilon\mu}$. We use this
observation to introduce an effective refraction index into the
equations for the scalar field.

Consider a new metric $dS_n^2$ related to (\ref{2.1}) as follows
\be
dS_n^2 = -\,\frac{A^2}{n}\,dt^2 + n\gamma_{ij}\,dx^i\,dx^j\,, \n{2.3}
\ee
where the effective refraction index $n=n(x^i)$ does not depend on time
$t$. We choose an equation for the scalar field $\varphi$ in the form
\be
\Box_n\varphi = 0\,, \n{2.4}
\ee
where $\Box_n$ is the ``box''-operator in the metric (\ref{2.3}).
Equation
\be
dS_n^2 = 0\,, \n{2.5}
\ee
gives characteristics for the equation (\ref{2.4}). Consider a Killing
observer moving with a four-velocity $u^{\mu}=\xi^{\mu}/A$. In the
reference frame of this observer, $d\tau = A\,dt$ is a proper time,
$dl=\sqrt{\gamma_{ij}\,dx^i\,dx^j}$ is a proper distance, and equation
(\ref{2.5}) takes the form
\be
\frac{dl}{d\tau} = \frac{1}{n}\,. \n{2.6}
\ee
The characteristics of the scalar field equation (\ref{2.4}) 
are rays moving with a velocity $c/n$ with respect
to a static observer, and hence $n$ really plays the role of the
refraction index.

The equation (\ref{2.4}) can be identically rewritten in the
form
\be
\Box\varphi = D(n)\,\varphi\,, \n{2.7}
\ee
where
\be
D(n)\,\varphi = \frac{n^2-1}{A^2}\,\partial_t^2\varphi \,. \n{2.8}
\ee
In the case when $|n-1|\ll 1$, the term $D(n)\varphi$ can be consided as
a perturbation\footnote{There is an ambiguity in the form of the metric
(\ref{2.3}). One can multiply the metric by any function  $f(n)$ which
for $n=1$ takes the value 1. This operation would modify the form of the
operator $D(n)$. In particular, a term proportional $\nabla n \nabla$ would be
generated. For a wave of characteristic frequency $\omega$ it gives a
contribution $\sim \omega \Delta n/b$ where $b$ is a size of the body.
It can be considered as a perturbation only if it is much smaller than
the leading derivative terms of the unperturbed operator which are of
the order $\omega^2$. For our problem $\omega\sim a$ where $a$ is the
acceleration of the body, and $ab\ll 1$. In order to escape problems
connected with the applicability of the perturbation approach and to
simplify the calculations we choose the special case $f=1$. 
}.
It should be emphasized that the form of the operator $D(n)$ implies
that equations (\ref{2.7})-(\ref{2.8}) can be easily generalized to the
case of a media with dispersion. For a monochromatic wave of frequency
$\omega$, $\partial_t^2\rightarrow -\omega^2$, and in order to take
into account the dispersion it is sufficient to put $n=n(\omega )$.

In our set-up we assume that $n=1$ everywhere outside some
four-dimensional region $\Gamma$ where a refractive body is located,
and $n=\,\mbox{const}\ \not= 1$ inside this region. We assume that a
body is static and rigid. Let $B$ be a three-dimensional volume occupied
by the body at $t=t_0$ and $S=\partial B$ be a surface of the body. 
Denote by $\Sigma$ a three-dimensional surface formed by
Killing trajectories passing through $S$. The region $\Gamma$ is defined
as a four-dimensional region located inside $\Sigma$. Thus we have
$\Sigma = \partial\Gamma$.

The operator (\ref{2.8}) in a spacetime with such a body is of the form
\be\n{2.9}
D(n) = \frac{n^2-1}{A^2}\,\vartheta (\Gamma )\,\partial_t^2\, ,
\ee
where $\vartheta$ is the Heavyside step function.

\subsection{Uniformly accelerated body}

Now we adopt the above scheme to the case of a uniformly accelerated
body moving in a flat spacetime. Let $(T,X,Y,Z)$ be standard Cartesian
coordinates so that the metric is
\be
dS^2 = -dT^2+dX^2+dY^2+dZ^2\,. \n{2.12}
\ee
Denote by $\gamma$ a world line of a uniformly moving observer. If $X$
is the direction of motion and $a$ is the acceleration, $\gamma$ is
described by the equation 
\be
X^2-T^2 = l^2 \equiv a^{-2}\,. \n{2.13}
\ee
In what follows the length parameter $l$ will play an important role.
For this reason it is convenient from the very beginning to introduce
dimensional coordinates
\be
t = \frac{T}{l}\,, \quad  x = \frac{X}{l}\,, \quad y = \frac{Y}{l}\,, 
\quad z = \frac{Z}{l}\,. \n{2.14}
\ee
Denote
\be
x = (1+\xi )\cosh\eta\,, \quad  t=(1+\xi )\sinh\eta\,. \n{2.15}
\ee
Then the metric (\ref{2.12}) takes the Rindler form
\be
dS^2 = l^2\left[-(1+\xi )^2\,d\eta^2 +d\xi^2 + dy^2 +dz^2\right]\,.
\n{2.16}
\ee
This metric is valid in the wedge $x>|t|$. The equation (\ref{2.13}) for
$\gamma$ takes the form
\be
\xi = 0\,, \n{2.17}
\ee
while $l\eta$ is a proper time along $\gamma$.

Surface $\eta = \eta_0$ is a plane with a three-dimensional flat metric.
It is convenient to introduce spherical coordinates $(\beta ,\theta
,\phi )$ related to $(\xi ,y,z)$ as
\be
(\xi ,y,z) = \beta n^i\,,\n{2.18}
\ee
where
\be
n^i = (\cos\theta ,\sin\theta\cos\phi , \sin\theta\sin\phi) \n{2.19}
\ee
is a unit vector directed from the origin $\xi = y = z =0$ to the point
$(\xi ,y,z)$.

Consider a small uniformly accelerated body, i.e. with the size much
smaller than $l$. Such a body is at rest in the reference frame
(\ref{2.16}). In the general case the surface of the body is described
by the equation 
\be
\beta = \beta_0(\theta,\varphi)\,. \n{2.20}
\ee
Relation (\ref{2.20}) is the equation which defines the surface
$\Sigma$, while (\ref{2.20}) together with $\eta = \eta_0$ determines a
position of the surface of the body at time $\eta_0$. Later we consider
a special case when the body is a ball and its surface is a sphere of
radius $b$. For this case
\be
\beta_0 = \frac{b}{l} =\mbox{const}\,. \n{2.21}
\ee

\section{Quantum Radiation of an Accelerated Refractive Body}\label{s3}
\setcounter{equation}0

\subsection{Hadamard functions and stress-energy tensor}
To find quantum radiation of an uniformly accelerated body with
$|n-1|\ll1$, we consider the operator $D(n)$ as a perturbation and use
perturbation expansion to obtain an answer.

In the absence of the body the field equation is
\be
\Box\varphi = 0\,. \n{3.1}
\ee
Denote
\be
G_0^{(1)}(x,x') = \la 0|\hat{\varphi}(x)\,\hat{\varphi}(x') + 
\hat{\varphi}(x')\,\hat{\varphi}(x)|0\ra \,, \n{3.2}
\ee
the Hadamard function for the standard Minkowski vacuum $|0\ra $. One has
\be
G_0^{(1)}(x,x') = \frac{1}{2\pi^2 l^2}\,\frac{1}{s^2(x,x')} \,, \n{3.3}
\ee
where
\be
s^2(x,x') = -(t-t')^2+(x-x')^2+(y-y')^2+(z-z')^2 \,, \n{3.4}
\ee
$(t,x,y,z)$ and $(t',x',y',z')$ being dimensionless Cartesian
coordinates of points $x$ and $x'$.

Consider the inhomogeneous equation
\be
\Box\varphi = j\,. \n{3.5}
\ee
Its solution is
\be
\varphi (x) = \varphi_0(x) -l^4\int G_0^{\ind{ret}}(x,x')\,j(x')\,d^4x'\,. 
\n{3.6}
\ee
where the retarded Green function $G_0^{\ind{ret}}$ is
\be
G_0^{\ind{ret}}(x,x') = \frac{1}{2\pi
l^2}\,\delta(s^2(x,x'))\,\vartheta(t-t')\,. \n{3.7}
\ee

By considering the right-hand side of (\ref{2.7}) as a perturbation and
using (\ref{3.6}) one gets
\be
G^{(1)}(x,x') = G_0^{(1)}(x,x') + G_{\ind{ren}}^{(1)}(x,x')\,, \n{3.8}
\ee
where
\[
G_{\ind{ren}}^{(1)}(x,x') = -l^4\,\left\{\int_{\Gamma}d^4x''\left[
G_0^{\ind{ret}}(x,x'')\,D''\,G_0^{(1)}(x,x'') \right.\right.
\]
\be
\hspace{4.5cm} \left.\left. +\,G_0^{\ind{ret}}(x',x'')\,D''\,G_0^{(1)}(x,x'')
\right]\right\}\,. \n{3.9}
\ee

Notation $D''$ indicates that the operator $D$ acts on the argument
$x''$. We also use notation $G_{\ind{ren}}^{(1)}$ for
$G^{(1)}-G_0^{(1)}$. That is this object that is required for the
calculation of physically observable quantities obtained by subtracting
contribution of zero-point fluctuations in an empty spacetime. The
integration in (\ref{3.9}) is performed over the interior of the
world-tube $\Gamma$.

By calculating $G_{\ind{ren}}^{(1)}(x,x')$ we can find $\la \varphi^2\ra
^{\ind{ren}}$ and $\la T_{\mu\nu}\ra ^{\ind{ren}}$ which characterize
change in the fluctuations and in the stress-energy tensor generated by
the moving body:
\be
\la \varphi^2(x)\ra ^{\ind{ren}} = \frac{1}{2}\,\lim_{x'\rightarrow x}
G_{\ind{ren}}^{(1)}(x,x')\,, \n{3.10}
\ee
\be
\la T_{\mu\nu}(x)\ra ^{\ind{ren}} = \lim_{x'\rightarrow x}\Pi_{\mu\nu'}\,
G_{\ind{ren}}^{(1)}(x,x')\,. \n{3.11}
\ee
Here
\be
\Pi_{\mu\nu'} = \left(\frac{1}{2}-\xi\right)\nabla_{(\mu}\nabla_{\nu')} +
\left(\xi -\frac{1}{4}\right)g_{\mu\nu'}\,g^{\alpha\beta}
\nabla_{\alpha}\nabla_{\beta'}
- \frac{1}{2}\,\xi\left(\nabla_{(\mu}\nabla_{\nu)} + 
\nabla_{(\mu'}\nabla_{\nu')}\right)\,, \n{3.12}
\ee
and $\xi$ is a parameter of non-minimal coupling. For $\xi =0$ one gets
the canonical stress-energy tensor, while for $\xi -1/6$ one gets the
``improved'' one with vanishing trace.

\subsection{$\la \varphi^2(x)\ra ^{\ind{ren}}$ and 
$\la T_{\mu\nu}(x)\ra ^{\ind{ren}}$ on ${\cal J}^+$}
We shall perform calculations assuming that a point of observation, $x$,
is located very far away from the moving body, in the so-called
radiation zone. Denote
\[
r=\sqrt{x^2+y^2+z^2}\,, \quad u=t-r\,, \quad (x,y,z) = rN^i\,, 
\]
\be
N^i = (\cos\Theta ,\sin\Theta\cos\Phi , \sin\Theta\sin\Phi )\,. \n{3.13}
\ee
The coordinate $u$ is  the dimensionless retarded time, and $(r,\Theta
,\Phi )$ are spherical coordinates in the inertial reference frame. The
wave-zone corresponds to taking the limit $r\rightarrow\infty$ with $u$
and $N^i$ fixed.

For calculations it is convenient to make the point splitting in
$t$-direction, i.e. to choose $x$ and $x'$ (the arguments of
$G_{\ind{ren}}^{(1)}$) so that
\be
r=r'\,, \quad N^i = N^{i'}\,. \n{3.14}
\ee
We also put
\be
u=u_0-\frac{h}{2}\,, \quad u'=u_0+\frac{h}{2}\,. \n{3.15}
\ee
For such a point splitting, $G_{\ind{ren}}^{(1)}(x,x')$ becomes a
function of $r,N^i,u_0$ and $h$:
\be
\left. G_{\ind{ren}}^{(1)}(x,x')\right|
_{{\ind{point} \atop \ind{splitted}} } = G(r,u_0,N^i,h) \,. \n{3.16}
\ee
Since $ G_{\ind{ren}}^{(1)}$ is a symmetric function of its coordinates
$x$ and $x'$, $G$ is an even function of its argument $h$.

To obtain $\la\varphi^2\ra ^{\ind{ren}}$, (\ref{3.10}), it is sufficient
to put $h=0$ in (\ref{3.16}). The calculations of the energy density
flux are more involved. One obtains
\be
\frac{dE}{dU d\Omega} \equiv   R^2\la T_{UU}\ra ^{\ind{ren}} 
\equiv   r^2\la T_{uu}\ra ^{\ind{ren}} =4\pi r^2
 \lim_{h\rightarrow
0}\left[\Pi_{uu}\,G(r,u_0,N^i,h)\right], \n{3.17}
\ee
where
\be
\Pi_{uu} = \frac{1}{8}\left(1-4\xi\right)\partial_{u_0}^2 -
\frac{1}{2}\,\partial_{h}^2\,. \n{3.18}
\ee
As we shall see, the leading term of $G$ in the wave zone is
proportional to $r^{-2}$, so that
\be
G(r,u_0,N^i,h) = \frac{1}{r^2l^2}\left[{\cal G}_1(u_0,\Theta) + 
h^2\,{\cal G}_2(u_0,\Theta)+O(h^4)\right]. \n{3.19}
\ee
We include factor $l^2$ to restore the correct dimensionality of $G$.
Functions ${\cal G}_1$ and ${\cal G}_2$ are dimensionless.  They do not
depend on $\Phi$ since the system is invariant with respect to rotation
in $(y,z)-$plane. Combining these results, we get
\be
\la\varphi^2\ra ^{\ind{ren}} \sim \frac{{\cal G}_1(u_0,\Theta)}
{2r^2l^2}\,, \n{3.20}
\ee
\be
\frac{dE}{dU d\Omega} \sim \frac{1}{l^2}\left[\frac{1}{8}\left(1-4\xi\right)
\partial_{u_0}^2{\cal G}_1(u_0,\Theta) -
{\cal G}_2(u_0,\Theta )\right] . \n{3.21}
\ee

\subsection{Boost invariance}

An important additional information about $\la\varphi^2\ra^{\ind{ren}}$
and the energy flux at ${\cal J}^+$ can be obtained by using the
symmetry of the problem. The spacetime with a uniformly accelerated body
is invariant under boost transformations
\[
t\rightarrow \tilde{t} = \gamma (t+vx)\,, 
\]
\be
x\rightarrow \tilde{x} = \gamma (x+vt)\,, \n {s.1}
\ee
\[
\tilde{y} = y\,, \quad \tilde{z} = z\,, \quad 
\gamma = (1-v^2)^{-1/2}\,. 
\]
It is easy to determine how this symmetry transformation acts on ${\cal
J}^+$. To do this, we introduce retarded spherical coordinates
$(u,r,\Theta ,\Phi )$ and
$(\tilde{u},\tilde{r},\tilde{\Theta},\tilde{\Phi})$ for both Minkowski
frames $x$ and $\tilde{x}$, and using (\ref{s.1}), we find the relation
between them. In the wave-zone limit, $r\rightarrow\infty $, $(u,\Theta
,\Phi )$ -- fixed, we have
\[
\tilde{r} = \gamma r(1+v\cos\Theta ) + \gamma\,\frac{vu\cos\Theta }
{1+v\cos\Theta }\,, 
\]
\be
\tilde{u} = \frac{\gamma^u}{1+v\cos\Theta }\,, \n {s.2}
\ee
\[
\tan\tilde{\Theta} = \frac{\sin\Theta}{\gamma (\cos\Theta +v)}\,, \quad 
\tilde{\Phi} = \Phi\,. 
\]
The invariance of $\la\varphi^2\ra^{\ind{ren}}$ under the transformation
(\ref{s.1}) requies that
\be
{\cal G}_1(\tilde{u},\tilde{\Theta}) = \gamma^2(1+v\cos\Theta )^2\,
{\cal G}_1(u,\Theta )\,. \n {s.3}
\ee
The invariance condition (\ref{s.3}) can be written in the infinitesimal
form. For this we put $v=\delta v$ and put the first variation of
(\ref{s.3}) with respect to $\delta v$ at $v=0$ equal to zero. This
gives the following relation
\be
\left(\frac{\partial\tilde{u}}{\partial v}\right)_{v=0}\partial_u{\cal
G}_1 + \left(\frac{\partial\tilde{\Theta}}{\partial v}\right)_{v=0}\partial_
{\Theta}{\cal G}_1 = 2\cos\Theta\,{\cal G}_1\,. \n {s.4}
\ee
Using (\ref{s.2}), we get
\be
\left(\frac{\partial\tilde{u}}{\partial v}\right)_{v=0} =
-u\cos\Theta\,, \quad \left(\frac{\partial\tilde{\Theta}}
 {\partial v}\right)_{v=0} = -\sin\Theta\,. \n {s.5}
\ee
Hence we have
\be
\frac{\partial{\cal G}_1}{\partial (\ln u)} + 
\frac{\partial{\cal G}_1}{\partial \ln (\sin {\Theta})} = -2{\cal G}_1\,. 
\n {s.6}
\ee
A general solution of the equation (\ref{s.6}) can be presented in the
form
\be
{\cal G}_1(u,\Theta ) = \frac{1}{u^2}\,{\cal H}_1(\frac{\sin\Theta }
{u})\,. \n {s.7}
\ee
In a similar way we get
\be
{\cal G}_2(u,\Theta ) = \frac{1}{u^4}\,{\cal H}_2(\frac{\sin\Theta }
{u})\,. \n {s.8}
\ee

\section{$\la\varphi^2\ra ^{\ind{ren}}$ and $\la T_{\mu\nu}\ra
^{\ind{ren}}$  in the Wave Zone }\label{s4}
\setcounter{equation}0

\subsection{Wave zone approximation}

To calculate functions ${\cal G}_1$ and ${\cal G}_2$, we rewrite
(\ref{3.9}) in a more explicit form. Since integration is performed
over the world tube $\Gamma$, it is convenient
to write the integral in (\ref{3.9}) in the Rindler coordinates
\be
\int_{\Gamma}d^4x'' \, \ldots = \int d\eta\int d\omega\int_0^{\beta_0(\theta ,
\phi)}\beta^2\, A\, d\beta \, \ldots \equiv \int d^4v \, \ldots \,, 
\n{4.1}
\ee
where $A=1+\beta\,\cos\theta$, $d\omega = \sin\theta\,d\theta\,d\phi$ 
and $\beta_0$ defines the
boundary of the body, see (\ref{3.20}). In these coordinates we also
have
\be
D = \frac{(n^2-1){\cal D}}{l^2} \,, 
\n{4.2}
\ee
\be
{\cal D} = A^{-2}\,\vartheta (\beta_0(\theta
,\phi ) - \beta )\,\,\partial_{\eta}^2\,. \n{4.3}
\ee

Both Green functions $G_0^{\ind{ret}}$ and $G_0^{(1)}$ depend only on the
distance $s$ between a point in the wave zone and a point inside or on
the boundary of the world tube $\Gamma$. Simple calculations give
\be
s^2(x,x'') = -2r\, w -u^2 + 2uA\sinh\eta +1+2\beta n^1 + 
\beta^2\,, \n{4.6}
\ee
where
\be
w =u +\beta \, \vec{n}_{\bot}\vec{N}_{\bot} -F(\eta)\, A\,, 
\hspace{0.5cm}
F(\eta)=\sinh\eta -N^1\cosh\eta \, . \n{4.7}
\ee
Here $(u,r,\Theta ,\Phi )$ are retarded spherical 
coordinates of the point $x$ in the wave
zone, and $(\eta ,\beta ,\theta ,\phi )$ are Rindler spherical 
coordinates of the point $x''$ in the tube $\Gamma$. The vectors
$\vec{n}$ and $\vec{N}$ are defined by equations (\ref{2.19}) and
(\ref{3.13}), respectively,  and
$\vec{n}_{\bot}=(0,n^2,n^3)$, $\vec{N}_{\bot}=(0,N^2,N^3)$.

An important observation is that the leading, $1/r^2$, term of
$G_{\ind{ren}}^{(1)}$ in the wave zone can be obtained by  neglecting 
the terms independent of $r$  in (\ref{4.6}), that is by using the
following approximate expressions
\be
s^2(x,x'') \approx -2rw_-\,, \quad s^2(x',x'') \approx -2rw_+\,, \n{4.8}
\ee
where
\be
w_{\pm} = w_0 \pm \frac{h}{2}\,, \quad w_0 = u_0  + \beta\,  
\vec{n}_{\bot}\vec{N}_{\bot}-F(\eta) \, A \,. \n{4.9}
\ee
Combining all these results and using (\ref{3.19}), we get
\[
{\cal G}_1(u_0,\Theta ) + h^2{\cal G}_2(u_0,\Theta ) + \ldots 
\]
\be
\hspace{1.1cm} = \frac{n^2-1}{16\pi^3}\int d^4 v\left[\delta
(w_-)\,{\cal D}(\frac{1}{w_+}) + \delta (w_+)\,{\cal D}(\frac{1}{w_-})
\right]. \n{4.10}
\ee
We omitted $\vartheta$-function which enters the definition (\ref{3.7})
of $G_0^{\ind{ret}}$ since a future-directed null cone emitted from a
point in the wave zone never crosses the tube $\Gamma$.

\subsection{Calculation of integrals}

Denote by $I(\eta_0,h)$ the following integral
\be
I(\eta_0,h) = \int_{-\infty}^{\infty}d\eta\,\delta(w_-)\,
\partial_{\eta}^2(\frac{1}{w_+})\,. \n{4.11}
\ee
Notice that
\be
\partial_{\eta}^2(\frac{1}{w_+}) = \frac{F \,A}{w_{\pm}^2} + 
\frac{2(F')^2\, A^2}{w_{\pm}^3}\,, \n{4.12}
\ee
where $(\ )'=\partial_{\eta}(\ )$. We use here the following property of
the function $F(\eta )$
\be
F''(\eta ) = F(\eta )\,. \n{4.13}
\ee
For $\Theta\not= 0$, $F(\eta )$ is a monotonically increasing function
which changes from $-\infty$ (at $\eta = -\infty$) to $+\infty$ 
(at $\eta = \infty$). Thus equation $w_0(\eta ) = c$ has a unique
solution for any $c$. We denote by $\eta_0$ and $\eta_{\pm}$ the
solutions of the following equations
\be
w_0(\eta_0) = 0\,, \quad w_0(\eta_{\pm}) = \mp\frac{h}{2}\,. \n{4.14}
\ee
The integral (\ref{4.11}) can be easily calculated by using the relation
\be
\delta(w_-) = \frac{\delta (\eta_-)}{F'(\eta_-)\,A}\,. \n{4.15}
\ee
We get
\be
I(\eta_0,h) = \frac{1}{F'(\eta_-)}\left[\frac{1}{h^2}F(\eta_-) + 
\frac{2}{h^3}(F'(\eta_-))^2 \,A\right]. \n{4.16}
\ee
Now we use the following Taylor expansions
\be
\eta_- =\eta_0 + \sum_{n=1}^{\infty}\frac{h^n}{n!}\,z_n\,, \n{4.17}
\ee
\be
w_0(\eta_-) = - A\,\sum_{n=1}^{\infty}\frac{1}{n!}\,F_0^{(n)}\,(\eta_- - 
\eta_0)^n\,, \n{4.18}
\ee
where $F_0^{(n)} = F^{(n)}(\eta_0)$. Substituting (\ref{4.17}) into
(\ref{4.18}) and using (\ref{4.13}), we solve equation
\be
w_0(\eta_-) = \frac{h}{2} \n{4.19}
\ee
to determine $z_n$ in terms of $F_0$ and $F'_0$. For calculations we use
Maple. The result is
\[
z_1 = -\frac{1}{2AF'_0}\,, \quad  z_2 = -\frac{F_0}{4A^2(F'_0)^3}\,, \quad 
z_3 = -\frac{-3(F_0)^2+(F'_0)^2}{8A^3(F'_0)^5}\,, 
\]
\be
z_4 = \frac{3F_0[-5(F_0)^2+3(F'_0)^2]}{16A^4(F'_0)^7}\,, \n{4.20}
\ee
\[
z_5 = -\frac{3[35(F_0)^4-30(F_0)^2(F'_0)^2+3(F'_0)^4]}{32A^5(F'_0)^9}\,.
\]
We performed calculations up to the order $h^5$ which is required to
calculate (\ref{4.11}) up to the order $h^2$.

Next steps of the calculations are the following:
\begin{enumerate}
\item Write $F(\eta_-)$ as
\be
F(\eta_-) = F_0\cosh (\eta_- -\eta_0) + F'_0\sinh (\eta_- -\eta_0)\,. \n{4.21}
\ee

\item Substitute (\ref{4.17}) into expression (\ref{4.21}) and using
(\ref{4.20}) get the expression for $F(\eta_-)$ in powers of $h$.
\item Use the obtained expansion to calculate $I(\eta_0,h)$ defined by
(\ref{4.16}).

\item Calculate $J_0(\eta_0,h)=I(\eta_0,h)+I_0(\eta_0,-h)$ to obtain the
expression for
\be
J(\eta_0,h) = \int_{-\infty}^{\infty}d\eta\left[\delta (w_-)\,
\partial_{\eta}^2(\frac{1}{w_+}) +\delta
(w_+)\,\partial_{\eta}^2(\frac{1}{w_-})\right]. \n{4.22}
\ee
\end{enumerate}

We performed these calculations using Maple. The final result is
\[
J(\eta_0,h) = -\frac{F_0\sin^2\Theta}{2A^2(F'_0)^5}
\left[ 1+
 \frac{h^2}{8A^2(F'_0)^4}\,[(4\cosh^2\eta_0+3)\cos^2\Theta
\right.
\]
\be
\quad\quad\quad\quad \left.
 -\,8\sinh\eta_0\cosh \eta_0\cos\Theta
-7+4\cosh^2\eta_0] \right] \,. \n{4.23}
\ee
Thus the right-hand side of (\ref{4.10}) is
\be\n{4.24}
{n^2-1\over 16\pi^3}\int d\omega \int_0^{\beta_0(\theta,\phi)} d\beta\,
\beta^2\, A^{-1}\, J(\eta_0,h)\, .
\ee
In this relation $\eta_0$ is a solution of the equation
\be\n{4.25}
F(\eta_0)=A^{-1} (u_0 +\beta\, \vec{n}_{\bot}\vec{N}_{\bot})\, .
\ee
Since $\beta\ll 1$, one can solve this equation perturbatively. Let
$\tilde{\eta}_0$ be a solution of the equation
\be\n{4.26}
F(\tilde{\eta}_0)=u_0\, ,
\ee
then 
\be\n{4.27}
\eta_0-\tilde{\eta}_0\approx {\beta\over {F'}_0}\,
(\vec{n}_{\bot}\vec{N}_{\bot} -u_0 \cos\theta)\, ,
\ee
where ${F'}_0={F'}(\tilde{\eta}_0)$. In order to take into account the
dependence of $J(\eta_0,h)$  on $\beta$ it is sufficient to expand $J$
near $\tilde{\eta}_0$ and use (\ref{4.27}). The obtained corrections
contain an additional small factor $b/l$, where $b$ is the size of the
body. By neglecting this correction and similar corrections in $A$, we get
\be\n{4.28}
{(n^2-1)\, V\over 16\pi^3 l^3} \,  J_0(\tilde{\eta}_0,h),
\ee
where  $V$ is the volume of the body. By comparison with
(\ref{4.10}) we get
\be\n{4.29}
{\cal G}_1(u_0,\Theta)= -{\cal B} \, 
\frac{F_0\sin^2\Theta}{(F'_0)^5}\, ,
\ee
\be
{\cal G}_2(u_0,\Theta)= -{\cal B} \, 
\frac{F_0\sin^2\Theta}{8(F'_0)^9}
\left(
4 F_0^2-3\sin^2(\Theta)
\right) \,,
\ee
where
\be\n{4.30}
{\cal B}={(n^2-1)\, V\over 32\pi^3 l^3} \, .
\ee
Let us emphasize once again that since we are considering the leading in
$b/l$ terms, to calculate $F_0$ and $F'_0$ one must put $b=0$. In
particular, $F_0=u_0$. In order to get $F'_0$ one must first solve the
equation
\be\n{4.30a}
F_0\equiv \sinh \eta_0 -\sin\Theta \cosh \eta_0 =u_0\, ,
\ee
and determine $\eta_0=\eta_0(u_0,\cos\Theta)$, ans substitute this value
into the definition of $F'_0$
\be\n{4.30b}
F'_0\equiv \cosh \eta_0 -\sin\Theta \sinh \eta_0 \, .
\ee

\subsection{$\la\varphi^2\ra ^{\ind{ren}}$ and energy density flux}

\begin{figure}
\centerline{\epsfig{file=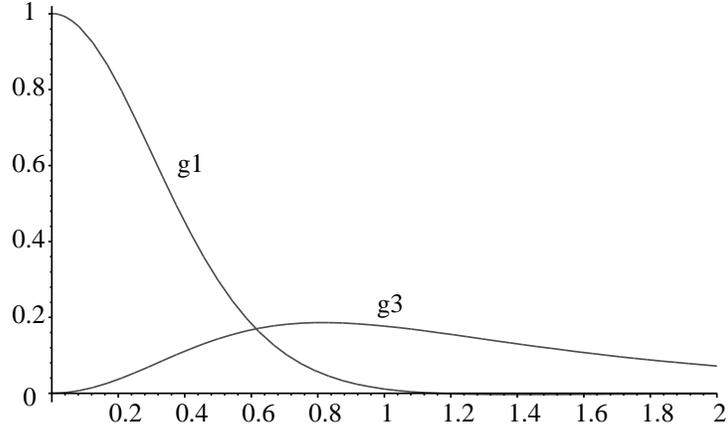, width=10cm}}
\caption[g]
{Plots of functions $g_1(z)$ ($g1$) and $g_3(z)$ ($g3$).} 
\label{g}
\end{figure}

It is easy to check that ${\cal G}_{1,2}$ obey the symmetry relations
(\ref{s.7})--(\ref{s.8}). Really by using relations (\ref{4.30a}) and
(\ref{4.30b}) one can rewrite ${\cal G}_{1,2}$ in the form
\be\n{4.31}
{\cal G}_1(u,\Theta)=-{\cal B}\, {g_1(z)\over u^2}\, ,
\hspace{0.5cm}
{\cal G}_2(u,\Theta)=-{\cal B}\, {g_2(z)\over u^4}\, ,
\ee
where $z=\sin(\Theta)/ u$ and
\be\n{4.32}
g_1(z)={z^2\over (1+z^2)^{5/2}}\, ,
\hspace{0.5cm}
g_2(z)={z^2 (4-3z^2)\over 8(1+z^2)^{9/2}}\, .
\ee
Using equations (\ref{3.20}) and (\ref{3.21}), we get
\be\n{4.33}
\la\varphi^2\ra ^{\ind{ren}} \sim -\frac{(n^2-1) b^3 a g_1(z)}
{48 \pi^2 R^2 U^2}\,, 
\ee
\be\n{4.34}
\frac{dE}{dU d\Omega} = -\frac{(n^2-1) b^3 a}{12\pi^2 U^4}
\, (1-5\xi)g_3(z) \,. 
\ee
Here
\be\n{4.35}
g_3(z)=\frac{1-3z^2/4}{(1+z^2)^{9/2}} \,. 
\ee
In the previous relations we restored dimensional coordinates $R$ and
$U=T-R$, and $z$ which enters these relations is $\sin(\Theta)/(a U)$,
where $a$ is the acceleration of the body.
Plots of functions $g_1$ and $g_3$ are shown in Figure~\ref{g}

By integrating over angles one gets the total energy density flux
\be\n{4.36}
\frac{dE}{dU} = -\frac{(n^2-1) b^3 a^3}{30\pi U^2}
\, (1-5\xi) f(aU) \,, 
\ee
where
\be\n{4.37}
f(u)={ 1+4u^2+5u^4+10u^6 \over (1+u^2)^4}\, .
\ee

\section{Conclusion}

In this paper we considered the vacuum polarization effects in the
presence of a uniformly accelerated dielectic body. We considered a scalar
field model. Under assumption that a size of the body, $b$, is much
smaller than the inverse acceleration, $a^{-1}$, we calculated the
energy density flux created by the accelerated body at infinity. This
flux is given by (\ref{4.36}), and for the canonical energy (i.e. for
$\xi=0$) it is always negative. Its divergence $\sim U^{-2}$ near $U=0$
is connected with the idealization of the problem: it is assumed that
the motion remains uniformly accelerated for an infinite interval of
time. The boost-invariance property connected with this assumption
significantly simplifies calculations. In particular, the quantity
$dE/(dU d\Omega)$ giving the anglular distribution of the energy density
flux at ${\cal J}^+$ at given moment of the retarted time $U$, besides
common scale dependence, $U^{-4}$, depends on the function of one
variable, $\sin\Theta/aU$. 

It would be interesting to repeat calculations for a more
realistic case of the electromagnetic field. One can expect that some
general features, especially those that are related to the symmetry of
the problem, will remain similar to the case of the scalar massless
field, while details, such as the angular distribution of the energy
flux which might depend on the spin of the field will differ. It should
be specially emphasized that the dependence of the electromagnetic 
field equations on the dielectric and magnetic properties of the media
is uniquely fixed, while in the model case of a scalar field there is
an ambiguity which we specially fixed to simplify the model. 

In our consideration we assumed that a uniformly accelerated refractive
body is cold, that is its temperature is zero. One can also consider a
case when a body is heated. Especialy interesting is a case when the
temperature of he body coincides with the Unruh temperature
corresponding to its acceleration. General arguments given in
\cite{ScCaDe:81,CaSc:83}
allows one to expect that in this case the quantum radiation vanishes.
We hope to return to these problems somewhere else.

\bigskip

\vspace{12pt}
{\bf Acknowledgments}:\ \  This work was  partly supported  by  the
Natural Sciences and Engineering Research Council of Canada. One of the
authors (V.F.) is grateful to the Killam Trust for its financial
support.

\end{document}